\documentclass[11pt]{article}
\usepackage{amsmath,amstext,amsbsy,amssymb}
\usepackage{bm}
\newcommand{\dso}{\Delta_{\scriptscriptstyle{ SO}}}
\newcommand{\vF}{v_{\scriptscriptstyle{ F}}}

\newcommand{\ssG}{{\scriptscriptstyle{ G}}}
\newcommand{\ssR}{{\scriptscriptstyle{ R}}}
\newcommand{\ssH}{{\scriptscriptstyle{ H}}}
\newcommand{\ssS}{{\scriptscriptstyle{ S}}}
\newcommand{\ssF}{{\scriptscriptstyle{ F}}}
\newcommand{\ssE}{{\scriptscriptstyle{ E}}}
\newcommand{\ssB}{{\scriptscriptstyle{ B}}}
\newcommand{\ssL}{{\scriptscriptstyle{ L}}}

\long\def\symbolfootnote[#1]#2{\begingroup%
\def\thefootnote{\fnsymbol{footnote}}\footnote[#1]{#2}\endgroup}

\begin{document}

\begin{center}

{\Large \bf Gauge Potential Formulations of the Spin Hall Effect in Graphene}

\vspace{2cm}

{\bf\"{O}mer F. Dayi}$^{a,b}$
\symbolfootnote[0]{{\it E-mail addresses:} dayi@itu.edu.tr and
dayi@gursey.gov.tr , yunt@itu.edu.tr } \,{and}\, {\bf Elif Yunt$^{a}$ }\\
\vspace{5mm}

{\em $^{a}${\it Physics Department, Faculty of Science and
Letters, Istanbul Technical University,\\
TR-34469, Maslak--Istanbul, Turkey} }

{\em $^{b}${\it Feza G\"{u}rsey Institute, P.O. Box 6, TR-34684,
\c{C}engelk\"{o}y, Istanbul, Turkey } }

\vspace{4cm}

{\bf Abstract}

\end{center}

Two different gauge potential methods are engaged to calculate explicitly the spin Hall conductivity in graphene.
The  graphene Hamiltonian with spin-orbit interaction is expressed in terms of kinematic momenta by introducing a gauge potential. 
A formulation of the spin Hall conductivity is  established by requiring that the time evolution of this kinematic momentum vector vanishes.  
We then  calculated  the conductivity employing the Berry gauge fields. We show that 
both of the gauge fields can be deduced from the pure gauge field arising from the Foldy-Wouthuysen transformations.

\vspace{1.5cm}


\renewcommand{\theequation}{\thesection.\arabic{equation}}
\newpage

\section{Introduction}

For low energies and at the Dirac points where conduction and
valance bands meet,  graphene is described by some copies of the
two-dimensional massless, free Dirac Hamiltonian
after substituting  the Fermi velocity $ v_{\ssF}$ for the
velocity of light $c$~\cite{semenoff}. 
Mass gaps can be generated by taking into account  spin-orbit interactions.
One of the interesting features of this  Dirac-like theory  
is the accomplishment of the spin Hall effect in graphene\cite{km}.
It is based on the observation that within this Dirac-like theory 
 the quantum Hall effect occurs in the absence of
external magnetic fields\cite{hal1}. This discovery of Kane and
Mele (KM)\cite{km} had far reaching consequences like a new
phenomena called topological insulators (see \cite{hk} for a
complete list of references). We would like to consider the  Dirac-like
Hamiltonian of  KM and calculate the spin Hall conductivity in
graphene directly using quantum mechanical and semiclassical methods.

We will present two different but somehow related approaches. The
first method relies on the observation that the 
Hamiltonian proposed by KM can be rephrased by defining a non-Abelian gauge
potential. We will deal mainly with the spin-orbit
interaction which preserves spin. When  the Hamiltonian is
expressed in terms of a gauge potential one can introduce the related kinematic
momentum vector and define its time evolution   which can be conceived as
the quantum mechanical analogue of  the force. 
In the case of the Hall  effect this is the  Lorentz force. One may provide  
a method of calculating the Hall conductivity  through the configurations obtained by requiring that the Lorentz force   vanishes
which  can be generalized to the  integer quantum Hall effect.  
We will show that acquainted  with the gauge potential   it is possible 
to derive explicitly the spin Hall conductivity in graphene imitating this
formulation. It  is in accord
with the result of KM.

The Foldy-Wouthuysen transformations of the two-dimensional massive Dirac Hamiltonian can be employed to define a non-Abelian gauge field\cite{bm}.
Because of being pure gauge field it leads to a vanishing field strength. However, in the adiabatic limit it generates the Abelian
 Berry gauge potential\cite{berry}. 
Examining the semiclassical equations of motion of electrons in the presence of Berry gauge fields,
one can show that the Hall conductivity is related to the Berry curvature.
We generalize this formulation to the Dirac-like Hamiltonian of KM and show explicitly that spin Hall conductivity
can be written in terms of the Berry phases. On the other hand
we will show that 
 the gauge potential  of the KM Hamiltonian which is furnished in the first method 
can be deduced from  this non-Abelian gauge field in the limit of vanishing momenta.

The method of calculating the  Hall conductivity through the configurations defined by the requirement that the time evolution of the  canonical momentum vector 
vanishes is presented in  Section \ref{iqh}.  In Section \ref{gr}
we first express  the KM Hamiltonian in terms of a vector field. We then imitate the formulation of  Section \ref{iqh} 
to derive the spin Hall effect conductivity in graphene  by taking into account  only  the spin conserving spin-orbit interaction. 
Section \ref{hbr} 
is devoted to the calculation of the spin Hall effect in graphene in terms of  the Berry phases.
In the last section we discuss the relation between these two different gauge fields and comment on whether the other possible interaction terms
can be treated similarly.

\setcounter{equation}{0}
\section{Integer quantum Hall effect \label{iqh} }

To illustrate the formalism which we will adopt to calculate the spin Hall conductivity in graphene we first
would like to discuss the integer quantum Hall effect.
Hall effect is a phenomena which occurs when
 electrons are constrained to move in a two-dimensional sample which is subjected to the uniform, perpendicular magnetic field  $B.$
Because of the Lorentz force the charge carriers  will move to one
edge of the sample creating a potential difference between the two
edges. By applying the in-plane, constant electric field $\bm E$
one can balance this potential difference thus the electrons will
move through the sample without deflection. Considering the
velocity of the electrons in this system one can find the Hall
current and derive the Hall conductivity\cite{gir}. It is possible
to carry this approach into quantum mechanics and obtain the
integer quantum Hall effect. To this aim let us introduce the
kinematic momenta $\pi_{i}=p_{i}-a_{i}$ associated with the
electromagnetic vector potential components 
$a_{i}=(eB/2c)\epsilon_{ij}r_j;\ i,j=1,2,$ 
and consider the
Hamiltonian
\begin{equation}\label{Hamiltonian}
    H_\ssB =\frac{\bm \pi^{2}}{2m} -e\bm E\cdot \bm r.
\end{equation}
The commutation relations obeyed by the
kinematic momentum and position operators are
$$
  [r_{i},\pi_{j}]=i\hbar\delta_{ij},
  [r_{i},r_{j}]  = 0,
  [\pi_{i},\pi_{j}]=i\hbar f_{ij},
$$
where $f_{ij}=\partial a_j/ \partial r_i-\partial a_i/ \partial r_j$ is the electromagnetic field strength with
the nonvanishing components
$$
f_{12}=-f_{21}=\frac{e}{c}B .
$$
The Heisenberg equations of motion following from the Hamiltonian
(\ref{Hamiltonian})  are
\begin{eqnarray}
\dot{r}_{j} &=& \frac{\pi_{j}}{m}, \label{Eom1} \\
    \dot{\pi}_{i}&=& eE_{i}+f_{ij}\dot{r}_{j} . \label{Eom2}
\end{eqnarray}
Obviously, the eigenvalues of  these operators correspond to the
velocity of charge carriers and the electromagnetic  force
acting on them, respectively. Thus quantum
mechanically  demanding that the electrons do not be
deflected by the electromagnetic force,  is equivalent to consider the states $\psi_\ssH
(\bm r )$ satisfying
\begin{equation}
\label{vef}
\dot{\bm \pi} \psi_\ssH (\bm r )=0.
\end{equation}
One can observe that nontrivial
$\psi_\ssH (\bm r )$ exists. This
ensures that we can consider the  subspace where  (\ref{vef}) is satisfied and solve
them to obtain the velocities of charge carriers which are not deflected as
\[
  v_i= \frac{c}{B}\epsilon_{ij}E_{j}.
 \]
Although we do not indicate explicitly these are the values of the ``velocity operators" (\ref{Eom1}) in
the subspace where the conditions (\ref{vef}) are fulfilled.
Plugging these into the definition of  the electrical current
$$
\bm{j}=-e\kappa\bm{v}
$$
where  $\kappa$ is the concentration of charge carriers, yields
$$
j_i=\sigma_{\ssH }\epsilon_{ij}E_j.
$$
Thus the Hall conductivity is obtained as
\begin{equation}\label{hallcondctvty}
    \sigma_{\ssH }=-\frac{ec\kappa}{B}.
\end{equation}

We should also specify $\kappa .$
The eigenstates of the Hamiltonian (\ref{Hamiltonian}) for $\bm E =0,$ which describes  the electrons moving on the plane
in the presence of the perpendicular, constant magnetic field $B$  are known as Landau levels.
They correspond to the eigenvalues
$$
    E_{\rm{n}}=\hbar \omega_{c}({\rm{n}}+\frac{1}{2});\ {\rm{n}}=1,2,\cdots ,
$$
where  $\omega_{c}= \frac{eB}{mc}$.
The density of states
of each
Landau level can be calculated to be
$$
    \rho_\ssL (E)=\frac{m}{2\pi\hbar^{2}}.
$$
For the two-dimensional free electrons one finds the same  density of states.
One can attain the concentration of charge carriers between the the ground state $E_0$ and the Fermi level $E^{(\ssL )}_\ssF$  as
\begin{equation}\label{nlandau}
    \kappa=\int_{E_0}^{E^{(\ssL )}_\ssF}\!\!\!\!\rho_\ssL (E) dE=\frac{m}{2\pi\hbar^{2}}(E^{(\ssL )}_\ssF -E_0).
\end{equation}
For Landau levels
$E_0=\frac{\hbar \omega_{c}}{2}$ and  $E^{(\ssL )}_\ssF=\hbar
\omega_{c}(N+\frac{1}{2})$ with $N$ being the index of the highest filled Landau
level, so that (\ref{nlandau}) yields
\begin{equation}
\kappa=\frac{eBN}{hc} .
\label{lakap}
\end{equation}
Inserting (\ref{lakap}) into (\ref{hallcondctvty}) one obtains
$$
    \sigma_{\ssH}=-\frac{e^{2}}{h}N,
$$
which characterizes the integer quantum Hall effect.

\setcounter{equation}{0}

\section{A gauge field formulation of  graphene \label{gr}}

Graphene is the one atom thick material composed  of carbon atoms which are situated at the corners of
hexagons arranged as a honeycomb
lattice. It is constructed as a superposition of two triangular sublattices which are
designated as $A$ and $B$. Its conduction and valance bands touch at  two inequivalent Dirac points
$K$ and $K^\prime .$  Around these  points, for low energies, the particles are described
by the free, massless, two-dimensional  Dirac-like Hamiltonian\cite{semenoff}
\begin{equation}\label{H0}
    H_{0}= v_{\ssF}\bm{\alpha}\cdot\bm{p},
\end{equation}
where $\vF $ is the effective velocity of electrons and 
\begin{equation}\label{alpha}
\bm{\alpha}\equiv \left(\sigma_{x}\tau_{z},\sigma_{y}\right).
\end{equation} 
$\sigma_{x,y,z}$ are the Pauli matrices acting
on the states of the sublattices $A$ and $B,$ in the representation where $\sigma_{z}={\rm diag} (1,-1).$
The other Pauli matrix $\tau_{z}={\rm diag} (1,-1) $ labels states at  the Dirac points $K$ and
$K^\prime .$ 
One can also introduce the Pauli matrices $s_{x,y,z}$ 
to identify the spin of electrons  with $s_{z}={\rm diag} (1,-1).$ 
We do not explicitly state neither the
direct product between these  spaces nor the  unit matrices. 
In the presence of the  uniform magnetic field  $B$ in $z$ direction  
the   spin-orbit interactions   proposed in \cite{km} yield the Hamiltonian 
\begin{equation}
\label{Hgeneral}
H^{\ssG}    =   H_{0}+\dso\sigma_{z}\tau_{z}s_{z}+\lambda_{\ssR}(\sigma_{x}\tau_{z}s_{y}-\sigma_{y}s_{x})-\frac{eB}{2}(y\sigma_{x}-x\sigma_{y}\tau_{z}).
\end{equation}
To imitate the formalism of Section \ref{iqh} we need to express the interaction Hamiltonian (\ref{Hgeneral}) by means of gauge fields.
Obviously, the magnetic field term can be written in terms of an electromagnetic gauge field. However, 
the fully fledged interaction Hamiltonian can be written as
$$
H^{\ssG} = \vF\bm{\alpha}\cdot\bm{\Pi}^{\ssG}\
$$
by substituting $\bm p$ in the free Hamiltonian (\ref{H0}) with the kinematic  
momentum vector
$$
\bm \Pi^{\ssG}=\bm p-\bm A^{\ssG},
$$
by introducing the following gauge potential components 
\begin{eqnarray}
    A_{x}^{\ssG}&=& i\frac{\dso}{2\vF }\sigma_{y}s_{z}-\frac{\lambda_{\ssR}}{\vF }s_{y}+\frac{eB}{2\vF } \tau_{z} y,\label{Ageneral1}\\
    A_{y}^{\ssG}&=&
    -i\frac{\dso}{2\vF }\sigma_{x}\tau_{z}s_{z}+\frac{\lambda_{\ssR}}{\vF }s_{x}-\frac{eB}{2\vF } \tau_{z} x. \label{Ageneral}
\end{eqnarray}
The last terms are the ordinary electromagnetic gauge fields in the symmetric gauge and the second terms can be read directly from (\ref{Hgeneral}).
The first terms are unusual and they arise through the algebra of the Pauli  matrices: $\sigma_{x}\sigma_{y}=i\sigma_{z}.$ 

The operators $\Pi_{i}^{\ssG}$ and $r_{i}$ satisfy the 
commutation relations
\begin{eqnarray}\label{ComRel}
    [r_{i},\Pi_{j}^{\ssG}] = i\hbar\delta_{ij},\ [r_{i},r_{j}] =0,\
    [\Pi_{i}^{\ssG},\Pi_{j}^{\ssG}]=i\hbar \epsilon_{ij} F_{xy}^{\ssG}.
\end{eqnarray}
The field strength is defined  as
$$
    F_{ij}^{\ssG}=\frac{\partial A_{j}^{\ssG}}{\partial r_{i}}-\frac{\partial A_{i}^{\ssG}}{\partial
    r_{j}}-\frac{i}{\hbar}[A_{i}^{\ssG},A_{j}^{\ssG}] .
$$
Thus, for the non-Abelian gauge field
(\ref{Ageneral1}), (\ref{Ageneral}),  
it takes the form $F^{\ssG}_{ij}=\epsilon_{ij}F^{\ssG}_{xy}$ with
\begin{equation}\label{Fxygeneral}
    F_{xy}^{\ssG}=\frac{eB}{\vF }\tau_{z}-\frac{\dso^{2}}{2\hbar \vF ^{2}}\sigma_{z}\tau_{z}
+\frac{i\dso\lambda_{\ssR}}{\hbar \vF ^{2}}(\sigma_{y}s_{y}+\sigma_{x}\tau_{z}s_{x})+\frac{2\lambda_{\ssR}^{2}}{\hbar \vF ^{2}}s_{z} .
\end{equation}

\subsection{Spin Hall effect in graphene \label{subs}}

To discuss  spin Hall effect we require that only the the spin-orbit term conserving the third component of spin is nonvanishing in (\ref{Hgeneral}). 
Hence, we deal with the Hamiltonian
\begin{equation}\label{H}
H=H_0 +\dso\sigma_{z}\tau_{z}s_{z},
\end{equation}
which can written as
$$
H=\vF\bm{\alpha}\cdot\bm{\Pi} ,
$$
through the kinematic momenta  $\bm{\Pi}=\bm{p}-\bm{A},$ where the gauge field $\bm{A}$ is  obtained
from (\ref{Ageneral1})-(\ref{Ageneral}) by setting $B=0,\ \lambda_{\ssR}=0:$
\begin{equation}\label{A}
    A_{x}= i\frac{\dso}{2\vF }\sigma_{y}s_{z},\ \ \
    A_{y}=-i\frac{\dso}{2\vF }\sigma_{x}\tau_{z}s_{z} .
\end{equation}
This is a vector potential taking values in  the $SU(2)$ group generated by  $\sigma_{x,y,z}.$
The curvature corresponding to (\ref{A}) is
$$
F_{xy}=\frac{-i}{\hbar}[A_x,A_y]=-\frac{\dso^{2}}{2\hbar \vF ^{2}}\sigma_{z}\tau_{z}.
$$
We are now equipped with the whole machinery needed to  
discuss the spin Hall effect  imitating the formulation of  the Hall conductivity presented in Section 2.
Considering the electrons in the external constant
electric field $\bm E$ described by 
\begin{equation}\label{H2}
H_\ssE=H-e\bm{E}\cdot\bm{r} ,
\end{equation}
we will
derive the Heisenberg equations of motion and obtain the
spin current by requiring that the time evolution of the kinematic momentum vector
vanishes.
The  ``velocity'' and ``force'' operators obtained through the Heisenberg equations of
motion $ \dot{\bm r}  =\frac{i}{\hbar}[ H_\ssE,\bm r ]$ and  $ \dot{\bm \Pi} = \frac{i}{\hbar}[ H_\ssE,\bm \Pi] $
 are
\begin{eqnarray}
    \dot{r}_{i}  &=& (\dot{x},\dot{y})=(\vF \sigma_{x}\tau_{z},\vF \sigma_{y}), \label{EO3-1}  \\
    \dot{\Pi}_{i} &=& \frac{i\dso}{\hbar}\sigma_{z}\tau_{z}s_{z}\Pi_{i}
-\frac{\dso^{2}}{2\hbar \vF ^{2}} \epsilon_{ij}\dot{r}_j\sigma_{z}\tau_{z}+eE_{i}.\label{EO3-2} 
\end{eqnarray}
Then, we demand that the ``force'' vanishes:
\begin{equation}
\dot{\bm \Pi} \Psi (\bm r )=0 .
\label{p0}
\end{equation}
One can easily  show that 
for all of the components of the spinor wave function $\Psi (\bm r )$
the same differential equation follows from  (\ref{p0}):
\begin{eqnarray}
\Psi^T (\bm r) =f (\bm r ) (1,1,1,1,1,1,1,1), \label{cosub1} \\
    \left[\frac{\dso^{2}}{\hbar^{2}}p_{i}^{2}+(eE_{i})^{2}+\frac{\dso^{4}}{\hbar^{2}\vF ^{2}}\right] f (\bm r )=0, \label{cosub2}
\end{eqnarray}
where $i=1,2$ is not summed over.
In the rest of this section we deal with operators taking values in the subspace spanned with the spinors (\ref{cosub1}) and (\ref{cosub2}),
although we will not explicitly write. In fact in this subspace we can now
solve (\ref{p0}) to obtain the velocities
\begin{equation}
\dot{r}_{i} = \frac{2i\vF ^{2}}{\dso}\varepsilon_{ij} \Pi_j s_{z}+ \dot{r}_{\ssH i},
\label{v}
\end{equation}
where we separated the electric field dependent part as
\begin{equation}
\label{ev}
\dot{r}_{\ssH i} = -\frac{2e\hbar
\vF ^{2}\varepsilon_{ij}E_{j}}{\dso^{2}}\sigma_{z}\tau_{z}.
 \end{equation}
We would like to calculate the spin Hall conductivity, therefore 
the relevant parts in the solution (\ref{v}) are the terms
proportional to the electrical field components (\ref{ev}).  Let us label the velocity of
spin up carriers by $\dot{r}_{\ssH  i}^{\uparrow}$ and the velocity of
spin down carriers by $\dot{r}_{\ssH i}^{\downarrow}$.
 We can further determine the velocities of particles in the $K$ and $K^\prime$
 valleys corresponding to $1$ and $-1$ eigenvalues of $\tau_{z}$ as
\begin{align}
\dot{r}_{\ssH i}^{\uparrow+} &=-\frac{2e\hbar \vF ^{2}
\varepsilon_{ij}E_{j}}{\dso^2}\sigma_{z},
   &
\dot{r}_{\ssH i}^{\uparrow-} &=\frac{2e\hbar \vF ^{2}
\varepsilon_{ij}E_{j}}{\dso^2}\sigma_{z} \label{yup+-},
\\
\dot{r}_{\ssH i}^{\downarrow+} &=-\frac{2e\hbar \vF ^{2}
\varepsilon_{ij}E_{j}}{\dso^2}\sigma_{z},
   &
\dot{r}_{\ssH i}^{\downarrow-}&=\frac{2e\hbar \vF ^{2}
\varepsilon_{ij}E_{j}}{\dso^2}\sigma_{z} . \label{ydown+-}
\end{align}
The Hall currents of the spin-up and spin-down electrons are defined as
\begin{eqnarray}
    {\bm j}^{\uparrow}_\ssH &=& n^{\uparrow +}{\dot{\bm r }}_{\ssH }^{\uparrow +}+n^{\uparrow-}{\dot{\bm r}}_{\ssH }^{\uparrow-} ,\label{hc1} \\
           \bm{j}^{\downarrow}_\ssH   &=&
               n^{\downarrow+}\dot{\bm{r}}_{\ssH }^{\downarrow+}+n^{\downarrow-}\dot{\bm{r}}_{\ssH }^{\downarrow-} , \label{hc2}
\end{eqnarray}
where 
$n^{\uparrow+},n^{\uparrow-},n^{\downarrow+},n^{\downarrow-}$
indicate  concentrations of the related carriers,
The Hall currents (\ref{hc1}), (\ref{hc2}) can be employed to define the spin Hall current as 
\begin{equation}\label{spincurrent}
\bm{j}_\ssH^\ssS  =\frac{\hbar}{2}(\bm{j}^{\uparrow}_\ssH -\bm{j}^{\downarrow}_\ssH ) .
 \end{equation}
We need to determine the concentrations
$n^{\uparrow+},n^{\uparrow-},n^{\downarrow+},n^{\downarrow-}.$
This will be elaborated inspecting  the corresponding Hamiltonians. In fact, 
there are four  different two-dimensional  Hamiltonians stemming from (\ref{H}):
$$
H=\left(
\begin{array}{cccc}
H^{\uparrow +} & 0&0 &0  \\
0& H^{\uparrow -}& 0& 0 \\
0& 0& H^{\downarrow +} & 0 \\
0 & 0& 0 & H^{\downarrow -}
\end{array}
\right).
$$
These two-dimensional  Hamiltonians
corresponding to the
$\uparrow,\ \downarrow$ spin and the $ K,\ K^\prime$ valley are
\begin{align}
H^{\uparrow+}
&=\vF (\sigma_{x}p_{x}+\sigma_{y}p_{y})+\dso\sigma_{z},
   &
H^{\uparrow-}
&=\vF (-\sigma_{x}p_{x}+\sigma_{y}p_{y})-\dso\sigma_{z},
 \label{Hup+-}
\\
H^{\downarrow+}
&=\vF (\sigma_{x}p_{x}+\sigma_{y}p_{y})-\dso\sigma_{z},
   &
H^{\downarrow-}&=\vF (-\sigma_{x}p_{x}+\sigma_{y}p_{y})+\dso\sigma_{z}.
 \label{Hdown+-}
\end{align}
The effect of the spin-orbit term  is to create a gap in the energy band structure of
the Hamiltonians. In terms of the eigenvalues of the momenta $\hbar \bm k$, (\ref{Hup+-}) and  (\ref{Hdown+-}) yield the same energy distribution
\begin{equation}\label{EnergySO}
    E= \pm \sqrt{\vF ^{2}\hbar^{2}k^{2}+\dso^{2}},
\end{equation}
corresponding to particle  and antiparticle (hole) states. 
We choose $\dso> 0$ and let the Fermi energy of graphene be in the gap by setting $\bm k =0 .$
Identifying the concentration of particles and  antiparticles (holes) by $n_{p}$ and $n_{a}$, we obtain 
\begin{align}
n^{\uparrow+} &= \left(%
\begin{array}{cc}
  n_{p} & 0 \\
  0 & n_{a} \\
\end{array}%
\right),
   &
n^{\uparrow-} &= \left(%
\begin{array}{cc}
  n_{a} & 0 \\
  0 & n_{p} \\
\end{array}%
\right),
 \label{nup+-}
\\
n^{\downarrow+} &= \left(%
\begin{array}{cc}
  n_{a} & 0 \\
  0 & n_{p} \\
\end{array}%
\right),
   &
n^{\downarrow-}&= \left(%
\begin{array}{cc}
  n_{p} & 0 \\
  0 & n_{a} \\
\end{array}%
\right).
 \label{ndown+-}
\end{align}
The derivations of (\ref{nup+-}) and (\ref{ndown+-}) are
elaborated in Appendix A. Inserting the carrier concentrations (\ref{nup+-})
and (\ref{ndown+-}) into the definition of the spin Hall current
(\ref{spincurrent}) one obtains
\begin{equation}\label{SCy1}
j^{\ssS}_{\ssH i}=-\frac{2e\hbar^{2}\vF ^{2}}{\dso^{2}}n\varepsilon_{ij}E_{j},
\end{equation}
where $n\equiv n_{p}-n_{a}.$  Now, $n$ can be calculated
using 
the density of states corresponding to the energy distribution (\ref{EnergySO})  found as
$$
    \rho_\ssE (E)=\int\frac{d^{2}k'}{(2\pi)^{2}}\delta(E-E')=\frac{|E|}{2\pi \vF ^{2}\hbar^{2}}.
$$
The edges of band gap lie at $\dso$ and
$-\dso ,$ hence when we consider positive  Fermi energy which lies in the gap, the number of states is derived as 
\begin{equation}\label{numbconc}
n=\int^{\dso}_0 \rho_\ssE (E)dE =\frac{|\dso|\dso}{4\pi\hbar^{2}\vF ^{2}}.
\end{equation}
Plugging  (\ref{numbconc}) into (\ref{SCy1}) yields
$$
j^\ssS_{\ssH i}=-\frac{e}{2\pi}\varepsilon_{ij}E_{j}
$$
and we obtain the spin Hall conductivity as
$$
    \sigma^{\ssS }_{\ssH }=-\frac{e}{2\pi}.
$$
This is in accord with the results of \cite{hal1,km}.

\setcounter{equation}{0}

\section{Spin Hall conductivity as Berry phase \label{hbr} }

Topological nature of the  Hall effect is well exhibited in terms of  Berry phases.
The semiclassical equations of motion are altered drastically in the presence of Berry gauge fields (see \cite{omer} and the references therein).
They yield an anomalous velocity term for electrons which leads to the
anomalous Hall conductivity. In fact, ignoring spin of electrons the Hall conductivity can be written in terms of
the Berry
curvature $\emph{F}_\ssB$ on the Fermi surface\cite{hal2, xsn} as (a complete list of references for the  Berry phase effects in
this context can be found in the recent review \cite{xcn}) 
\begin{equation}\label{3_1}
    \sigma_{\ssH }=-\frac{e^2}{\hbar}\int^{E_\ssF}\frac{d^{2}p}{(2\pi\hbar)^{2}}\emph{F}_\ssB .
\end{equation}
Considering the electrons with spin, a generalization of (\ref{3_1}) to 
the spin Hall effect was discussed in  \cite{ccn}.
Note that
in this section $\bm p$ is not a quantum operator but denotes the classical phase space variable.
We deal with four different two-dimensional Dirac-like theories (\ref{Hup+-}), (\ref{Hdown+-}),
 thus we should take into account the contributions arising from each of them separately.
We adopt the formulation of \cite{bm} to derive the Berry gauge fields arising from 
each one of the two-dimensional  Hamiltonians 
(\ref{Hup+-}), (\ref{Hdown+-}).
Therefore, we should start with giving  the unitary Foldy-Wouthuysen transformations $(U^{\uparrow +},U^{\uparrow -},U^{\downarrow +}, U^{\downarrow -})$
corresponding to the Dirac-like Hamiltonians $(H^{\uparrow +},H^{\uparrow -},H^{\downarrow +}, H^{\downarrow -}).$
We would like to present them in the unified notation: 
$$
U\equiv {\rm diag} (U^{\uparrow +},U^{\uparrow -},U^{\downarrow +}, U^{\downarrow -}).
$$
The unitary
Foldy-Wouthuysen transformation $U$ can be engaged to define the
 gauge field\cite{mnz, bm} 
$$
    \bm{{\cal A}}\equiv {\rm diag} ({\cal A}^{\uparrow +},{\cal A}^{\uparrow -},{\cal A}^{\downarrow +}, {\cal A}^{\downarrow -})
=i\hbar U(\bm{p})\frac{\partial U^\dag(\bm{p})}{\partial \bm{p}}.
$$

Exploring the Dirac-like Hamiltonians (\ref{Hup+-}), (\ref{Hdown+-}), we introduce the following Foldy-Wouthuysen transformation
\begin{equation}
U=\frac{1}{\sqrt{2E(E+\dso)}}\left(
\begin{array}{cccc}
\sigma_zH^{\uparrow +}+E & 0&0 &0  \\
0& -\sigma_zH^{\uparrow -}+E& 0& 0 \\
0& 0& -\sigma_zH^{\downarrow +}+E & 0 \\
0 & 0& 0 & \sigma_zH^{\downarrow -}+E
\end{array}
\right) ,\label{MU}
\end{equation}
where $E$ is the positive energy depending on $\bm p$ as
$
E=  \sqrt{\vF ^{2}\bm p^{2}+\dso^{2}}. 
$
Observe that (\ref{MU})  is defined to satisfy
\begin{equation}
\label{diag}
 UHU^\dagger =E\sigma_z\tau_z s_z.
\end{equation}
One can study 
each entry of (\ref{MU})  as in \cite{bm} and show that they lead to the gauge potential
\begin{eqnarray}
    \bm{{\cal A}}= \frac{i\hbar }{2E^{2}(E+\dso)}\left[ \vF E(E+\dso)\bm{\alpha}\sigma_z\tau_z s_z+ \vF ^{3}\sigma_z\tau_z s_z
    (\bm{\alpha}\cdot\bm{p})\bm{p}  + 
    \vF ^{2}E(\bm{\alpha}\cdot
    \bm{p})\bm{\alpha}-\vF ^{2}\bm{p}E\right] .\label{gaugefw}
\end{eqnarray}
Its components can be written explicitly as
\begin{eqnarray}
  {\cal A}_{x} &=&\frac{ \hbar( \vF E(E+\dso)\sigma_{y}s_z- \vF ^{3}(\sigma_{y} p_{x}-\sigma_{x}\tau_z p_{y})s_z p_{x}+ \vF ^{2}
E\sigma_{z}\tau_{z}p_{y})}{2E^{2}(E+\dso)} ,\label{gaugefwx}\\
  {\cal A}_{y} &=&\frac{\hbar(-\vF E(E+\dso)\sigma_{x}\tau_z s_z- \vF ^{3}(\sigma_{y} p_{x}- \sigma_{x}\tau_z p_{y})s_z p_{y}-
 \vF ^{2} E\sigma_{z}\tau_{z}p_{x})}{2E^{2}(E+\dso)}  . \label{gaugefwy}
\end{eqnarray}
Because of being a pure gauge potential 
the field strength of (\ref{gaugefw}) vanishes.
However, one can consider the adiabatic approximation 
by projecting on the positive energy states:
\begin{equation}
\label{gaugeproj}
   \bm{{\cal A}}^{B} \equiv \textsl{P} \bm{{\cal A}} \textsl{P}.
\end{equation}
By inspecting the positive eigenvalues of the (\ref{diag}) one can deduce that the projection operator $\textsl{P}$ is
$$
\textsl{P} \equiv {\rm diag} (\textsl{P}^{\uparrow}_{+},\textsl{P}^{\uparrow}_{-},\textsl{P}^{\downarrow}_{+},\textsl{P}^{\downarrow}_{-})= {\rm diag} (1,0,0,1,0,1,1,0).
$$
It should be noted that when projected
on positive energy states only the last terms in (\ref{gaugefwx})
and (\ref{gaugefwy}) make nonvanishing contributions, so that 
the Abelian Berry gauge field is
\begin{equation}
\label{bgfi}
 {\cal A}^\ssB_{i}=\frac{\hbar \vF ^{2}}{2E(E+\dso)}\epsilon_{ij} p_j {\bm 1}_\tau s_z  ,
\end{equation}
where  the unit matrix ${\bm 1}_\tau$ in the $\tau_z$ space is exhibited explicitly. 
It is worth  emphasizing that in $\bm{{\cal A}}$ negative energy states are present, thus it possesses twice the matrix elements
of  $\bm{{\cal A}}^\ssB .$ 
The nonvanishing component of the 
Berry curvature  is given as
\begin{equation}\label{Fberry}
{\cal F} =\hbar\left(\frac{\partial {\cal A}^\ssB_y}{\partial p_x}- \frac{\partial {\cal A}^\ssB_x}{\partial p_y}\right)\equiv {\rm diag} ({\cal F}^{\uparrow}_{+},
{\cal F}^{\uparrow}_- ,{\cal F}^{\downarrow}_{+} , {\cal F}^{\downarrow}_{-}) = -\frac{\hbar^2\vF^2 \dso}{2E^3} {\bm 1}_\tau s_z  .
\end{equation}

We propose to generalize (\ref{3_1}) to the spin Hall effect in graphene as 
\begin{equation}
\label{bpc}
    \sigma^\ssS_{\ssH }=-\frac{e}{2}\int^{E^{(2)}_\ssF}\frac{d^{2}p}{(2\pi\hbar)^{2}}
    \left[ ({\cal F}^{\uparrow}_{+}
+{\cal F}^{\uparrow}_-)-({\cal F}^{\downarrow}_{+} +{\cal F}^{\downarrow}_{-} )\right].
\end{equation}
$E^{(2)}_\ssF$  denotes the highest energy level occupied in the two-dimensional system.
Thus, inserting (\ref{Fberry}) into the definition (\ref{bpc}) leads to
\begin{eqnarray}
  \sigma_\ssH^{\ssS  } &=& -\frac{e}{2}\int_{-\infty}^{E^{(2)}_\ssF}\frac{d^{2}p}{(2\pi \hbar)^{2}} \left(-\frac{2\hbar^2\vF^2 \dso}{E^3}\right) \nonumber \\
                  &=& -\frac{e}{2\pi}\frac{\dso}{E^{(2)}_\ssF}. \label{nsos}
\end{eqnarray}
This in accord with the calculation of the spin Hall conductivity obtained by employing the
 Kubo formula which is  presented in Appendix B.

We let the Fermi energy level of graphene 
lie in the gap, so that  in (\ref{nsos})  we set $E^{(2)}_\ssF=\dso$ and obtain the
spin Hall conductivity as
$$
    \sigma_{\ssH }^{\ssS }=-\frac{e}{2\pi} .
$$
This is the value established in \cite{km}.

\section{Discussions}

The main difference between the gauge fields (\ref{A}) and (\ref{bgfi}) is
the fact that the latter  is acquired from (\ref{gaugefw}) in the adiabatic limit (\ref{gaugeproj}), thus it is within the particle bands
of the two-dimensional Dirac-like theories, but  the former one
is  non--Abelian which connects the   particle and hole bands of the two-dimensional Dirac-like theories. 
However, also  the  non--Abelian gauge potential (\ref{A}) can be obtained from the gauge field  (\ref{gaugefw}) in the 
vanishing momentum limit, up to a constant , as
$$
\bm A =\left(\frac{i\dso^2}{\hbar \vF^2}\right)\bm{{\cal A}}|_{\bm p =0} .
$$
This limit  corresponds to restrict  the Fermi energy of graphene to lie in the gap. In fact, the 
calculation of the spin Hall conductivity presented in Section \ref{subs}  is valid only for the Fermi energy lying in the gap.
Therefore, although the calculation methods are different the gauge fields (\ref{A}) and (\ref{bgfi})
result from the same gauge field (\ref{gaugefw}) in two different limits.

The nonvanishing magnetic field $B\neq 0$  case can be studied  similarly. However 
if we switch on the Rashba term $\lambda_R\neq 0$ in (\ref{Hgeneral}), the third component of spin is not conserved  and in spite of the fact
that  we can introduce the gauge potential  (\ref{Ageneral1}),(\ref{Ageneral}), 
the related curvature (\ref{Fxygeneral}) is
not diagonal in the spin space. Hence the method of Section \ref{subs} is not anymore suitable to discuss the spin Hall effect.

\setcounter{equation}{0}

\setcounter{section}{1} \null

\renewcommand{\theequation}{\Alph{section}.\arabic{equation}}
\renewcommand{\thesection}{\Alph{section}}

\section*{Appendix A}
Introducing the label
$I=(\uparrow\!\!\!+, \uparrow\!\!\!- , \downarrow\!\!\!+, \downarrow\!\!\!- )$ 
corresponding to each of the
two-dimensional Dirac-like Hamiltonians given in (\ref{Hup+-}), (\ref{Hdown+-}),
we  define the number operators  
as
\begin{equation}\label{c1}
    N^I  =n_{p}|p^I\rangle\langle p^I|+n_{a}|a^I\rangle\langle a^I| ,
\end{equation}
where 
$|p^I\rangle$ and $|a^I\rangle$ denote, respectively, the positive energy $E=\sqrt{v_{\ssF}^2\hbar^2 k^2+\dso^2}$ and the negative energy $-E$ eigenspinors of the
related Dirac-like Hamiltonians. 
For instance,
let us consider the Hamiltonian for the
 spin up carriers in
the $K$ valley, $H^{\uparrow+}$, 
 whose eigenspinors can be written in the chiral basis as
\begin{eqnarray}
  |p^{\uparrow+}\rangle &=& \left(%
\begin{array}{c}
  \cos(\frac{\theta}{2}) \\
  \sin(\frac{\theta}{2})e^{i\phi} \\
\end{array}%
\right) ,\label{spinorup+p}\\
  |a^{\uparrow+}\rangle&=& \left(%
\begin{array}{c}
  \sin(\frac{\theta}{2}) \\
  -\cos(\frac{\theta}{2}) e^{i\phi} \\
\end{array}%
\right),\label{spinorup+a}
\end{eqnarray}
where $\cos \theta =\frac{\dso}{E}$
and $\tan \phi=\frac{k_{y}}{k_{x}}$. Plugging them into (\ref{c1}) leads to
\begin{equation}\label{c2}
     N^{\uparrow+} ({\bm k})=\left(%
\begin{array}{cc}
  \frac{n_{p}+n_{a}}{2}+\frac{n_{p}-n_{a}}{2}\cos \theta & \frac{n_{p}-n_{a}}{2}e^{-i\phi} \sin \theta  \\
  \frac{n_{p}-n_{a}}{2} e^{i\phi} \sin \theta  & \frac{n_{p}+n_{a}}{2}-\frac{n_{p}-n_{a}}{2}\cos \theta
\end{array}%
\right).
\end{equation}
When the Fermi level of graphene is in the gap generated by the spin-orbit
interaction we should set $\bm k =0,$ hence
$\cos \theta =1$ and (\ref{c2}) yields
$$
 N^{\uparrow+} ({\bm k}=0)\equiv n^{\uparrow+}=\left(%
\begin{array}{cc}
  n_{p} & 0 \\
  0 & n_{a} \\
\end{array}%
\right).
$$
In order to obtain the concentrations corresponding to the other Hamiltonians, 
we need to consider their eigenspinors. In fact,  for the spin up carriers in the
$K^\prime$ valley described with   $H^{\uparrow -}$ the eigenspinors are
\begin{eqnarray}
  |p^{\uparrow-}\rangle &=& \left(%
\begin{array}{c}
  \sin(\frac{\theta}{2}) \\
  -\cos(\frac{\theta}{2})e^{-i\phi} \\
\end{array}%
\right) ,\label{spinorup-p}\\
  |a^{\uparrow-}\rangle&=& \left(%
\begin{array}{c}
  \cos(\frac{\theta}{2}) \\
  \sin(\frac{\theta}{2}) e^{-i\phi} \\
\end{array}%
\right).\label{spinorup-a}
\end{eqnarray}
Inserting them into (\ref{c1}) and considering the vanishing momentum limit leads to
$$
N^{\uparrow-} ({\bm k}=0)= n^{\uparrow-},
$$
where $n^{\uparrow-}$ is given in (\ref{nup+-}).
Similarly, for the spin down carriers in the $K$ valley,  we can show that the eigenspinors of the Hamiltonian $H^{\downarrow+}$ are
\begin{eqnarray}
  |p^{\downarrow+}\rangle &=& \left(%
\begin{array}{c}
  \sin(\frac{\theta}{2}) \\
  \cos(\frac{\theta}{2})e^{i\phi} \\
\end{array}%
\right) ,\label{spinordown+p}\\
  |a^{\downarrow+}\rangle&=& \left(%
\begin{array}{c}
  \cos(\frac{\theta}{2}) \\
  -\sin(\frac{\theta}{2}) e^{i\phi} \\
\end{array}%
\right).\label{spinordown+a}
\end{eqnarray}
Making use of them in (\ref{c1}) and setting $\bm k=0$ 
yield the  concentration number $n^{\downarrow+}$  given in (\ref{ndown+-}):
$$
N^{\downarrow+} ({\bm k}=0)= n^{\downarrow+}.
$$
The eigenspinors of the Hamiltonian $H^{\downarrow-}$ corresponding to  the spin down carriers in the $K^\prime$ valley are
\begin{eqnarray}
  |p^{\downarrow-}\rangle &=& \left(%
\begin{array}{c}
  \cos(\frac{\theta}{2}) \\
  -\sin(\frac{\theta}{2})e^{-i\phi} \\
\end{array}%
\right) ,\label{spinordown-p}\\
  |a^{\downarrow-}\rangle&=& \left(%
\begin{array}{c}
  \sin(\frac{\theta}{2}) \\
  \cos(\frac{\theta}{2}) e^{-i\phi} \\
\end{array}%
\right).\label{spinordown-a}
\end{eqnarray}
One can easily observe that they result in
$$
N^{\downarrow-} ({\bm k}=0)= n^{\downarrow-},
$$
where  $n^{\downarrow-}$ is given in (\ref{ndown+-}).

\setcounter{equation}{0}\setcounter{section}{2} \null

\renewcommand{\theequation}{\Alph{section}.\arabic{equation}}
\renewcommand{\thesection}{\Alph{section}}

\section*{Appendix B}

Kubo formula  corresponding to the Hamiltonians  (\ref{Hup+-}), (\ref{Hdown+-}) can be written in the notation of Appendix A as\cite{tknn}
\begin{equation}
\label{Kubo}
\left(\sigma^{\ssS}_{\ssH}\right)^I=\frac{e\hbar^2}{2}\int^{E^{(2)}_\ssF}\frac{d^2k}{(2\pi)^2}
\frac{2 {\rm{Im}} \left[ \langle
a^I|\dot{y}|p^I\rangle\langle p^I|\dot{x}|a^I\rangle \right]}{4E^2} ,
\end{equation}
where $\dot{x},$ $\dot{y}$ are the related velocity operators which can be read from (\ref{EO3-1}) and 
$$E^{(2)}_\ssF=\sqrt{v_{\ssF}^2\hbar^2 k_\ssF^2+\dso^2}.$$
For $H^{\uparrow+}$  the velocity operators are
$\dot{x}=v_{\ssF}\sigma_x$, $\dot{y}=v_{\ssF}\sigma_y$ and  the
eigenspinors $|p^{\uparrow+}\rangle$ and
$|a^{\uparrow+}\rangle$ are given  in (\ref{spinorup+p}) and
(\ref{spinorup+a}). Employing them in (\ref{Kubo}) leads to\cite{sinitsyn}
\begin{equation}\label{kubosigmaup+}
(\sigma^{\ssS}_{\ssH})^{\uparrow+}=
-\frac{e}{8\pi}\frac{\dso}{E^{(2)}_\ssF} .
\end{equation}
For the spin up carriers in the
$K^\prime$ valley, we set
$\dot{x}=-v_{\ssF}\sigma_x$, $\dot{y}=v_{\ssF}\sigma_y$ and  deal with
the eigenspinors (\ref{spinorup-p}),(\ref{spinorup-a}). We obtain the same conductivity 
\begin{equation}\label{kubosigmaup-}
(\sigma^{\ssS}_{\ssH})^{\uparrow-}=-\frac{e}{8\pi}\frac{\dso}{E^{(2)}_\ssF} .
\end{equation}
The contributions arising  from the spin down carriers in the $K$ and $K^\prime$ valleys are  also equal
but differ in sign with the spin up contributions: 
\begin{equation}\label{kubosigmadown}
(\sigma^{\ssS}_{\ssH})^{\downarrow+}=(\sigma^{\ssS}_{\ssH})^{\downarrow-}=\frac{e}{8\pi}\frac{\dso}{E^{(2)}_\ssF}.
\end{equation}
To obtain the spin Hall conductivity we should take the difference of the spin up and spin down contributions as
\begin{equation}\label{kubosigmaspin}
\sigma^{\ssS}_{\ssH}=((\sigma^{\ssS}_{\ssH})^{\uparrow+}+(\sigma^{\ssS}_{\ssH})^{\uparrow-})-((\sigma^{\ssS}_{\ssH})^{\downarrow+}+(\sigma^{\ssS}_{\ssH})^{\downarrow-}).
\end{equation}
Inserting (\ref{kubosigmaup+}), (\ref{kubosigmaup-}) and (\ref{kubosigmadown}) into (\ref{kubosigmaspin}) leads to
the spin Hall conductivity
$$
\sigma^{\ssS}_{\ssH}=-\frac{e}{2\pi}\frac{\dso}{E^{(2)}_\ssF}.
$$
This is the same with the result obtained in terms of the Berry phase (\ref{nsos}).

\end{document}